%% file: main.tex
\documentclass[sigconf,10pt, nonacm]{acmart}
\usepackage[ruled,linesnumbered]{algorithm2e}
\usepackage{algorithmic}
\usepackage{multirow}
\usepackage{graphicx}
\usepackage{todonotes}
\usepackage{xcolor}
\usepackage{xspace}
\usepackage{amsmath}
\usepackage{enumitem}
\usepackage{soul}
\newcommand{\name}{MobiVital\xspace}

\AtBeginDocument{%
  \providecommand\BibTeX{{%
    \normalfont B\kern-0.5em{\scshape i\kern-0.25em b}\kern-0.8em\TeX}}}

\begin{document}


\title{\name: Self-supervised Time-series Quality Estimation for Contactless Respiration Monitoring Using UWB Radar}

\author{Ziqi Wang}
\authornote{The research reported in this paper was conducted at UCLA and Samsung Research America, and is unrelated to this author's current affiliation.}
    \affiliation{
    \institution{Qualcomm}
    \streetaddress{}
    \city{} 
    \state{} 
    \country{}
    \postcode{}
    }
\email{wangzq312@g.ucla.edu}

\author{Derek Hua}
\affiliation{\institution{UCLA}
    \streetaddress{}
    \city{} 
    \state{} 
    \country{}
    \postcode{}
}
\email{derekhua@ucla.edu}

\author{Wenjun Jiang}
\affiliation{\institution{Samsung Research America}
    \streetaddress{}
    \city{} 
    \state{} 
    \country{}
    \postcode{}
}
\email{wenjun.jiang@samsung.com}
  
\author{Tianwei Xing}
\authornote{The research was conducted when these authors were affiliated with Samsung Research America and is unrelated to these authors' current affiliations.}
\affiliation{\institution{Meta}
    \streetaddress{}
    \city{} 
    \state{} 
    \country{}
    \postcode{}
}
\email{twxing@ucla.edu}

\author{Xun Chen}
\authornotemark[2]
\affiliation{\institution{Independent Researcher}
    \streetaddress{}
    \city{} 
    \state{} 
    \country{}
    \postcode{}
}
\email{xunchen@outlook.com}

\author{Mani Srivastava}
\authornote{The author holds concurrent appointments as an Amazon Scholar and Professor at UCLA, but the work in this paper is not associated with Amazon.}
\affiliation{%
  \institution{UCLA and Amazon}
    \streetaddress{}
  \city{} 
  \state{} 
  \country{}
  \postcode{}
}
\email{mbs@ucla.edu}

\renewcommand{\shortauthors}{Ziqi Wang, et al.}

\begin{abstract} 

Respiration waveforms are increasingly recognized as important biomarkers, offering insights beyond simple respiration rates, such as detecting breathing irregularities for disease diagnosis or monitoring breath patterns to guide rehabilitation training.
Previous works in wireless respiration monitoring have primarily focused on estimating respiration rate, where the breath waveforms are often generated as a by-product. As a result, issues such as waveform deformation and phase inversion have largely been overlooked, reducing the signal's utility for applications requiring breathing waveforms. To address this problem, we present a novel approach, \name, that improves the quality of respiration waveforms obtained from ultra-wideband (UWB) radar data. \name combines a self-supervised autoregressive model for breathing waveform extraction with a biology-informed algorithm to detect and correct waveform inversions. To encourage reproducible research efforts for developing wireless vital signal monitoring systems, we also release a 12-person, 24-hour UWB radar vital signal dataset, with time-synchronized ground truth obtained from wearable sensors. Our results show that the respiration waveforms produced by our system exhibit a 7-34\% increase in fidelity to the ground truth compared to the baselines and can benefit downstream tasks such as respiration rate estimation. The code, checkpoints, and dataset of this paper can be found at \url{https://github.com/nesl/mobivital-public}.


\end{abstract}

\maketitle

\input{Contents/01-Introduction}

\input{Contents/02-Motivation}

\input{Contents/03-Dataset.tex}

\input{Contents/04-Method.tex}
\input{Contents/05-Results}

\input{Contents/06-RelatedWork}
\input{Contents/07-FinalSessions.tex}

\bibliographystyle{ACM-Reference-Format}
\bibliography{citations}

\end{document}

%% file: Contents/01-Introduction.tex
\section{\textbf{Introduction}}\label{Sec:intro}

\textbf{Motivation}. Respiration is a critical biomarker that reveals rich information about human body functionalities. Respiration rate has been a vital indicator used in healthcare and exercise to diagnose diseases, detect stress, and assess physical exertion~\cite{nicolo2020importance}. The \emph{respiration waveform}, which captures the detailed pattern of chest movement during breathing, is a valuable biomarker beyond simple respiration rate. Apart from the rate, respiration waveform provides insights into each breath's depth, rhythm, and variability~\cite{kunczik2022breathing, raji2019knitted}, which can reveal subtle physiological and pathological changes inside human bodies. For instance, irregularities in the waveform, such as asymmetrical or interrupted patterns, may indicate early signs of respiratory or cardiac dysfunction, as seen in conditions like sleep apnea~\cite{kang2020non}, chronic obstructive pulmonary disease (COPD)~\cite{colasanti2004analysis} or Parkinson's disease~\cite{yang2022artificial}. Furthermore, respiration waveform analysis can help improve the effectiveness of athletic training~\cite{neumann2009relationship} or monitor diet habits~\cite{dong2013wearable}. Also, rehabilitation training involving breath exercises can benefit from respiration waveform monitoring as it provides critical feedback to ensure patients follow the designed breath patterns~\cite{carmo2024magnetic, rahman2022breathebuddy}.

Wearable chest bands attached to the subject's chest wall can provide the most accurate breath waveform readings by directly measuring the chest expansion caused by respiration~\cite{raji2019knitted}. However, wearing such chest bands may cause discomfort and distractions and hinder body movement. Contactless methods are preferred due to their noninvasive nature. 
With the advancement of sensing technologies, wireless signals are employed as novel approaches for respiration monitoring. Typical sensing modalities include acoustic waves \cite{xu2019breathlistener}, Wi-Fi~\cite{liu2021wiphone,xie2024robust,zeng2020multisense}, RFID~\cite{yang2020respiration, zang2022rfid}, mmWave~\cite{yang2016monitoring, hao2024mmwave}, and Impulse-Radio Ultra-Wideband (IR-UWB) radar~\cite{zheng2020v2ifi,chen2021movi,xie2022passive}. These systems can measure the subject's vital signals from a distance with minimal discomfort and constraints by employing wireless signals. Among all these modalities, we focused on the IR-UWB Radar for its low power consumption, high resolution (can detect subtle vibrations), and ranging capability (can separate multiple subjects based on their distance).

\textbf{Current Solution.} An IR-UWB respiration sensing system can be summarized into a three-step procedure: (1) Preprocessing. In this stage, complex-valued UWB data matrice is converted to real values via magnitude~\cite{yang2019multi, zheng2020v2ifi} and phase calculation~\cite{gu2013hybrid,xie2023short}. Background removal algorithms are also applied to exclude clutters from static objects. (2) Waveform extraction. UWB radar produces two-dimensional radar matrices, one axis corresponds to object range (distance) and the other corresponds to time. The system needs to detect the distance of the human subject and slice the matrix on the range axis to extract a waveform as the respiration measurement. For target ranging, existing works explore algorithms such as signal power variance~\cite{husaini2022non, yang2019multi}, autocorrelation~\cite{shen2018respiration}, and constant false alarm rate (CFAR) detection~\cite{zheng2020v2ifi, zheng2021more}. Some other works leverage additional information from another sensor (e.g., a depth camera)~\cite{xie2022passive,wang2022capricorn}. (3) Application-specific processing. Filtering~\cite{kim2019non}, mode decomposition algorithms~\cite{husaini2022non, zheng2020v2ifi}, or wavelet analysis~\cite{husaini2022non} are combined with spectrum analysis to extract respiration rates. Literature also reported the usage of end-to-end deep-learning models in this stage~\cite{xie2022deepvs}.

\begin{figure}[thbp]
    \centering
    \vspace{-5pt}
    \includegraphics[width=\linewidth]{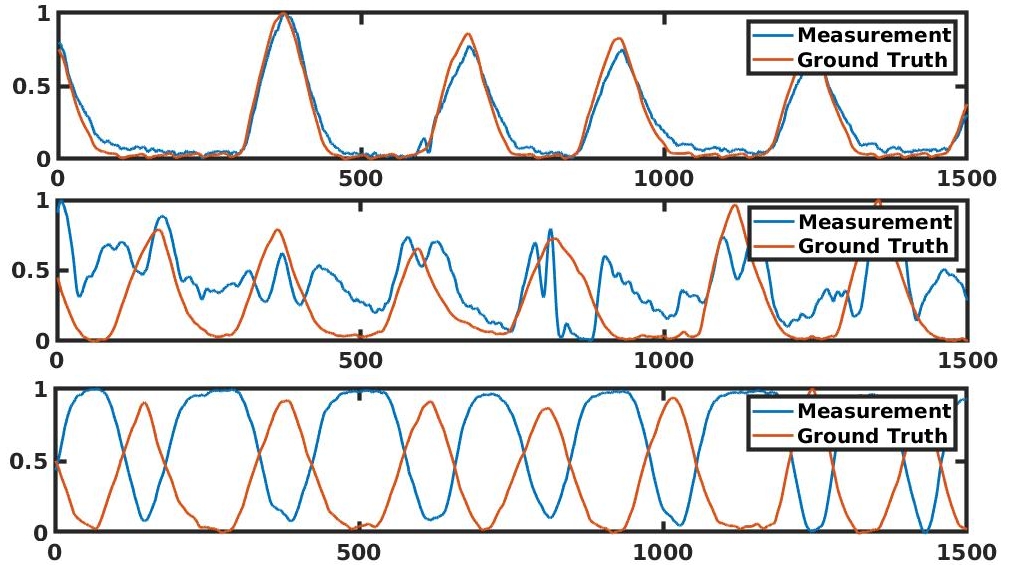}
    \vspace{-15pt}
    \caption{Example of good, distorted, and inverted UWB measurements.}
    \label{fig:bagsig}
\end{figure}

\textbf{Problem}. With a focus on estimating the respiration rate, existing works often produce respiration waveform as a by-product. Despite the importance of respiration waveform monitoring, breath waveform morphology issues, i.e., \emph{distortion} and \emph{inversion}, are often overlooked, as they have a limited impact on respiration rate estimation. In Figure~\ref{fig:bagsig}, we showcase these common signal quality issues in the breath waveform. We attribute these quality issues to suboptimal respiration waveform extraction from the UWB data matrix: human respiration typically affects a range of more than 50~\!cm in the UWB data matrix, which produces many possible candidate time series.
The time series selected based on the target distance estimation is often suboptimal. 
Through dataset collection and analysis, we discovered that a better candidate time series is often available from the same UWB data matrix. Picking the best candidate is trivial if the ground truth waveform is available: we can sequentially compare the candidate time series against the ground truth to pick the best non-inverted time series with minimum distortions. However, how can we predict the signal quality and detect signal inversion without leveraging the ground truth?

\textbf{Proposed Method}. To address the challenges above, we present the \name system, which includes: (1) a predictor of signal quality to assist in picking the best candidate time series, in order to reduce signal distortion and (2) a detector of signal inversion, without accessing the ground truth. \textbf{For signal quality estimation}, we propose a self-supervised procedure inspired by a pioneer work~\cite{xie2023short}. This procedure leverages the generalization limitation of machine learning models, turning it from a foe to a friend.
Machine learning models are known for performing well on data samples with a similar distribution as the training data, and generalizing poorly on out-of-distribution. Thus, we train an autoregressive model (using a chuck of the time series to predict its near future) with high-quality respiration data. During the deployment, we feed the candidate time series to this autoregressive model. Hypothetically, a good-quality data sample has a similar distribution as the training dataset and we can expect good prediction accuracy. On the other hand, poor-quality data are unseen in this model and we can expect a high prediction loss. Thus, the model prediction accuracy becomes a natural surrogate of the signal quality and requires no labels. \textbf{For inversion}, we develop a lightweight algorithm inspired by the biological nature of the respiration movement. During inhalation, chest, and abdominal muscles contract to draw air into the lungs. Upon exhalation, these muscles relax, allowing the lungs to deflate~\cite{national2022how}. Natural respiration behavior often exhibits a ``duty cycle'' of less than half, as inhalation requires active muscle effort. Thus, we can design a signal processing pipeline to calculate such a duty cycle and detect phase inversion.

We conduct a comprehensive evaluation of our proposed method. The results prove that the autoregressive model is a good surrogate for signal quality. The time series selected by the model has a high correlation with the ground truth compared with baseline methods and exhibits higher accuracy when used by a downstream model to estimate respiration rate. Through an ablation study, we also demonstrate the effectiveness of our phase inversion detector. We introduce the system design in detail in Section~\ref{Sec:design} and analyze the results in Section~\ref{Sec:results}. 

The development of wireless vital signal monitoring systems often requires the construction of complex signal processing pipelines~\cite{zheng2020v2ifi, zhang2021contactless, shyu2020uwb} or the training of deep neural networks~\cite{zheng2021more,chen2021movi,xie2022deepvs}, which leads to the demand for high-quality datasets with time-synchronized ground truth. Unfortunately, such datasets are rarely open-sourced as a community resource. To encourage reproducible research in this community, we release our 24-hour large-scale dataset collected from 12 subjects on vital signal sensing using IR-UWB radar.
We also provide ground truth waveforms collected from wearable sensors, which are time-synchronized with the radar time series. This study is IRB-approved and the anonymized dataset will be publicly released (IRB approval number and dataset link omitted for anonymity). We introduce more details of \name in Section~\ref{Sec:data}. In summary, our major contributions are outlined as follows:
\begin{itemize}
    \item A 12-person, 24-hour open-source dataset for respiration monitoring using UWB radar.
    \item A self-supervised autoregressive model to predict radar-measured respiration quality with no knowledge of the ground truth.
    \item A bio-informed lightweight algorithm to detect if a respiration signal is inverted.
\end{itemize}

%% file: Contents/02-Motivation.tex
\section{Preliminary Study}\label{sec:prelim}

\begin{figure}[h!]
    \centering
    \includegraphics[width=\linewidth]{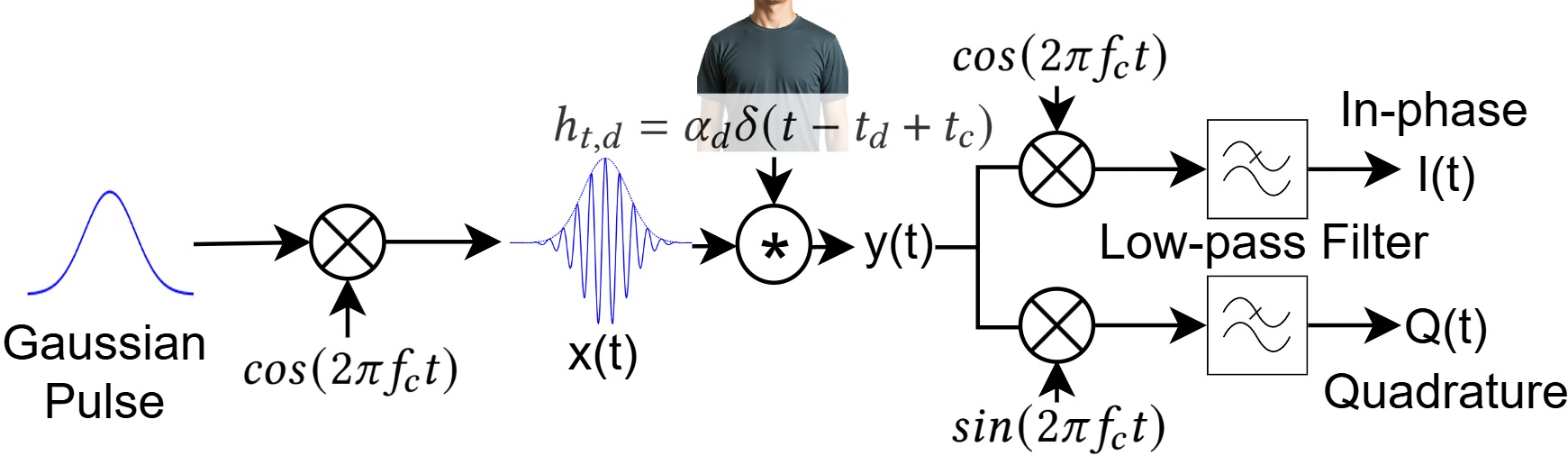}
    \caption{UWB respiration sensing theory.}
    \label{fig:theory}
\end{figure}

This section briefly overviews the theory of UWB-based vital signal sensing, the data structure, and the signal quality issues. We first start by introducing how respiration is sensed by the IR-UWB radar, as is shown in Figure~\ref{fig:theory}. 

The transmitter sends repetitive picosecond pulses $s(t)=Ae^{-\frac{x}{2\sigma^2}}$, which is then modulated by the carrier frequency $cos(2\pi f_c t)$. 
The environment is modeled as a multipath time delay channel $\sum_l \alpha_l \delta(t-t_l)$, where $\alpha_l$ and $t_l$ indicate path attenuation and delay. 
On the receiver end, it is possible to separate the paths based on their delays (lengths). If we focus on the path reflected on the human chest wall, the received signal can be represented as the transmitted signal $x(t)$ convolved with channel impulse response $h_{t,d}$, i.e.,
\begin{equation}
    y(t) = x(t) * h_{t,d} = \alpha_d  \delta(t-t_d+t_c).
\end{equation}
Here $t_d$ is decided by the distance between the subject and the sensor, and $t_c$ is a dynamic term describing the path length change caused by respirations.

On the receiver end, $y_t$ is converted to baseband by multiplying with the carrier frequency $cos(2\pi f_c t)$ and the phase-shifted carrier frequency $sin(2\pi f_c t)$, and then low-pass filtered to generate in-phase (I) and quadrature signal (Q), which can be represented as a complex number $w(t) = I+jQ$, where
\begin{equation}
\label{equ:rec}
    W(t) = s(t-t_d+t_c) e^{-j 2 \pi f_c t_d + j 2 \pi f_c t_c}.
\end{equation}
In other words, both the amplitude and phase of the received baseband signal $W(t)$ is a function of the respiration-induced delay $t_c$ and can be used for respiration sensing.

\begin{figure}[htb]
    \centering
    \includegraphics[width=0.98\linewidth]{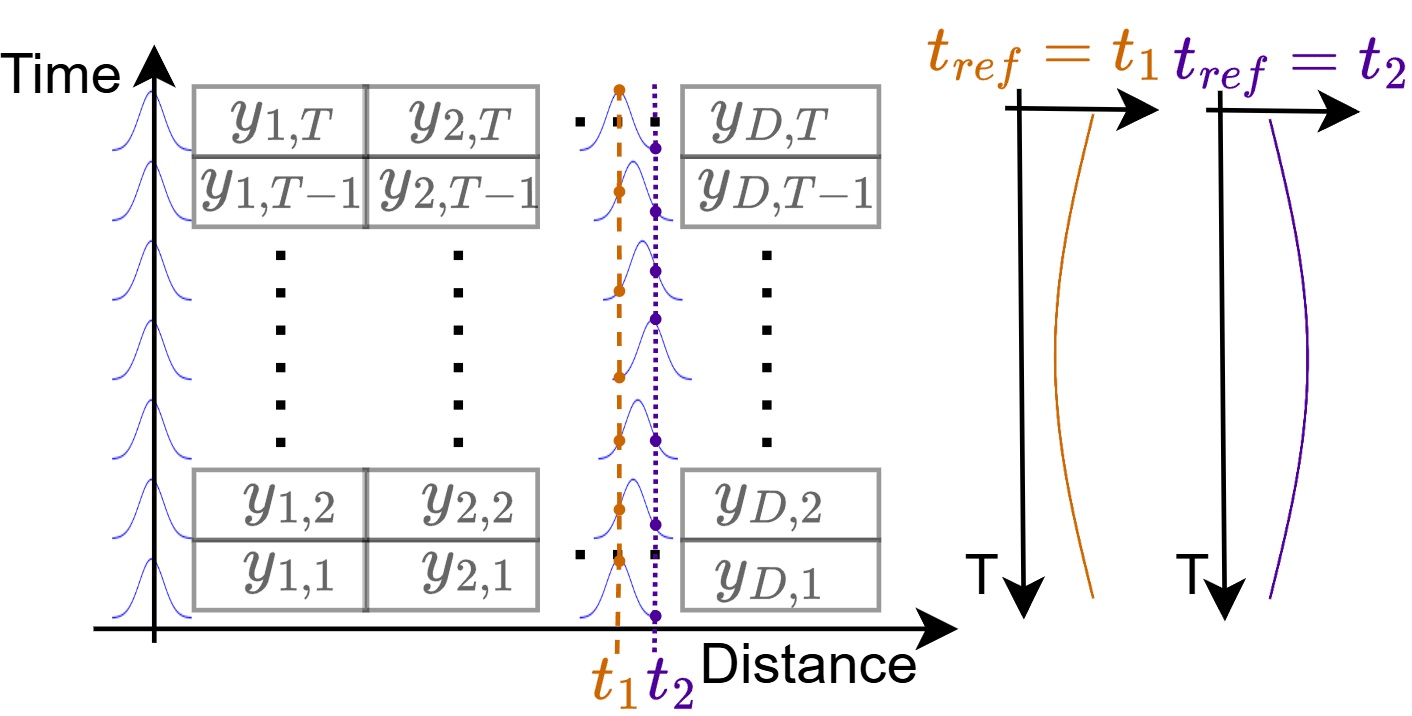}
    \vspace{-10pt}
    \caption{UWB data structure and a simple illustration of inversion.}
    \label{fig:data}
    \vspace{-15pt}
\end{figure}

Typically, a UWB radar produces a two-dimensional matrix $\mathbf{M} \in \mathrm{C}^{D\times T}$ (see Figure~\ref{fig:data}. One axis is time and the other is range (the distance from the sensor to the target objects, sometimes known as ``fast time''). Each value on this matrix is a complex number, as shown in Equ.~\ref{equ:rec}.
Our UWB radar sensor produces distance bins of 0.0514~\!m, which means the distance axis is discretized into 5~\!cm bins. As we have analyzed, UWB radars measure human breath using the chest-abdomen movement. 

To derive the breath signal, an important step is to take a slice along the range (distance) axis to ``pick'' a distance that best represents the human body. State-of-the-art systems either rely on the data matrix features such as variance~\cite{yan2016through} or range-FFT~\cite{zheng2020v2ifi}, or additional information from another sensor (e.g., a depth camera)~\cite{xie2022passive, wang2022capricorn} to determine the human presence at a particular distance. However, the human body has a complicated reflection surface which leads to multiple paths of different lengths, and thus respiration typically affects 5-10 adjacent distance bins. The range detected algorithms introduced above are often suboptimal for extracting respiration waveforms.
\begin{figure}[b]
    \centering
    \vspace{-10pt}
\includegraphics[width=\linewidth]{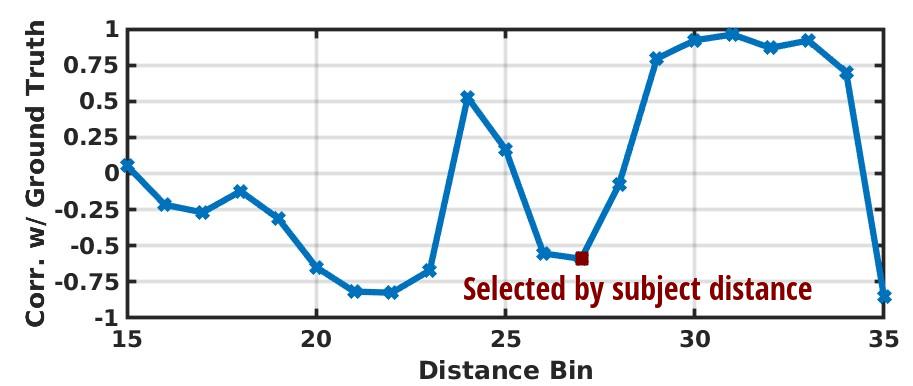}
    \vspace{-20pt}
    \caption{Respiration signal quality across different distance bins.}
    \label{fig:sigqual}
\end{figure}
We illustrate the signal quality issue in Figure~\ref{fig:sigqual}. The x-axis is the distance bin, and the y-axis is the correlation between the respiration waveform in that bin and the ground truth waveform. A close-to-one score indicates a high-quality signal. We can see that despite the optimal bin choice being 30-33, the human presence detector yields bin 27, whose signal contains a lot of noise and \textbf{deformation}. Also, at bins 26, 27, and 35, we observe a negative correlation score, which means the measured waveform moves in the opposite direction of the chest movement, i.e., \textbf{inversion}, which can also be observed in other research papers~\cite{higashikaturagi2008non, fallatah2022monitoring}. The cause of inversion can be complicated. Figure~\ref{fig:data} shows bin discretization as one of the possibilities. In the figure, we use blue pulses to indicate the pulses reflected on the human chest. The subject's chest moved away from the sensor and then back, indicating breathing out and then breathing in. If we slice the distance axis at fast time $t_1$ that corresponds to the location of the fully expanded chest, we see a normal waveform as indicated in orange. However, if the discrete bin happens to slice the signal at $t_2$ (fully collapsed chest), we see an inverted waveform (purple). Another possible reason for inversion is the subject's breath habit. Some people may exhibit chest breathing (shallow breathing where the belly squeezes as the chest expands) whereas others may use belly breathing (deep breathing where the stomach expands together with the chest).

Nevertheless, we have shown that (1) signal quality issues such as deformation and inversion are common in the UWB-measured breathing waveforms, and (2) a better choice of range bin is often available which provides time series of a higher quality. To address this problem, we propose a solution in this paper that can predict the candidate signal quality, without knowing the ground truth breathing waveform.

%% file: Contents/03-Dataset.tex
\section{\textbf{MobiVital Dataset}}\label{Sec:data}
We dive into the details of \name by first introducing its dataset. This section covers the data collection platform, the data collection protocol, and some dataset statistics. The collection of our \name is IRB-approved, and we will make the dataset publicly available to encourage reproducible research in the wireless vital signal sensing area.

\begin{table*}[htb]
\begin{tabular}{ccccccccccccc}
\hline
\multicolumn{1}{|c|}{Volunteer}                                                          & \multicolumn{1}{c|}{A}     & \multicolumn{1}{c|}{B}    & \multicolumn{1}{c|}{C}     & \multicolumn{1}{c|}{D}     & \multicolumn{1}{c|}{E}    & \multicolumn{1}{c|}{F}    & \multicolumn{1}{c|}{G}    & \multicolumn{1}{c|}{H}    & \multicolumn{1}{c|}{I}    & \multicolumn{1}{c|}{J}    & \multicolumn{1}{c|}{K}    & \multicolumn{1}{c|}{L}    \\ \hline
\multicolumn{1}{|c|}{\begin{tabular}[c]{@{}c@{}}Duration/mins\\ (Tripod)\end{tabular}}   & \multicolumn{1}{c|}{112.5} & \multicolumn{1}{c|}{79.5} & \multicolumn{1}{c|}{107.0} & \multicolumn{1}{c|}{104.0} & \multicolumn{1}{c|}{64.0} & \multicolumn{1}{c|}{55.0} & \multicolumn{1}{c|}{68.0} & \multicolumn{1}{c|}{71.5} & \multicolumn{1}{c|}{76.0} & \multicolumn{1}{c|}{64.0} & \multicolumn{1}{c|}{76.0} & \multicolumn{1}{c|}{59.5} \\ \hline
\multicolumn{1}{|c|}{\begin{tabular}[c]{@{}c@{}}Duration/mins\\ (Handheld)\end{tabular}} & \multicolumn{1}{c|}{94.5}  & \multicolumn{1}{c|}{27.0} & \multicolumn{1}{c|}{107.0} & \multicolumn{1}{c|}{98.5}  & \multicolumn{1}{c|}{32.0} & \multicolumn{1}{c|}{28.0} & \multicolumn{1}{c|}{32.0} & \multicolumn{1}{c|}{27.0} & \multicolumn{1}{c|}{32.0} & \multicolumn{1}{c|}{36.0} & \multicolumn{1}{c|}{32.0} & \multicolumn{1}{c|}{-}    \\ \hline
\multicolumn{1}{l}{}                                                                     & \multicolumn{1}{l}{}       & \multicolumn{1}{l}{}      & \multicolumn{1}{l}{}       & \multicolumn{1}{l}{}       & \multicolumn{1}{l}{}      & \multicolumn{1}{l}{}      & \multicolumn{1}{l}{}      & \multicolumn{1}{l}{}      & \multicolumn{1}{l}{}      & \multicolumn{1}{l}{}      & \multicolumn{1}{l}{}      & \multicolumn{1}{l}{}   
\end{tabular}
\vspace{-10pt}
\caption{Dataset Statistics.}\label{tab:stats}
\vspace{-20pt}
\end{table*}

\subsection{Data Collection Platform}
\begin{figure}[h!]
    \centering
    \includegraphics[width=0.9\linewidth]{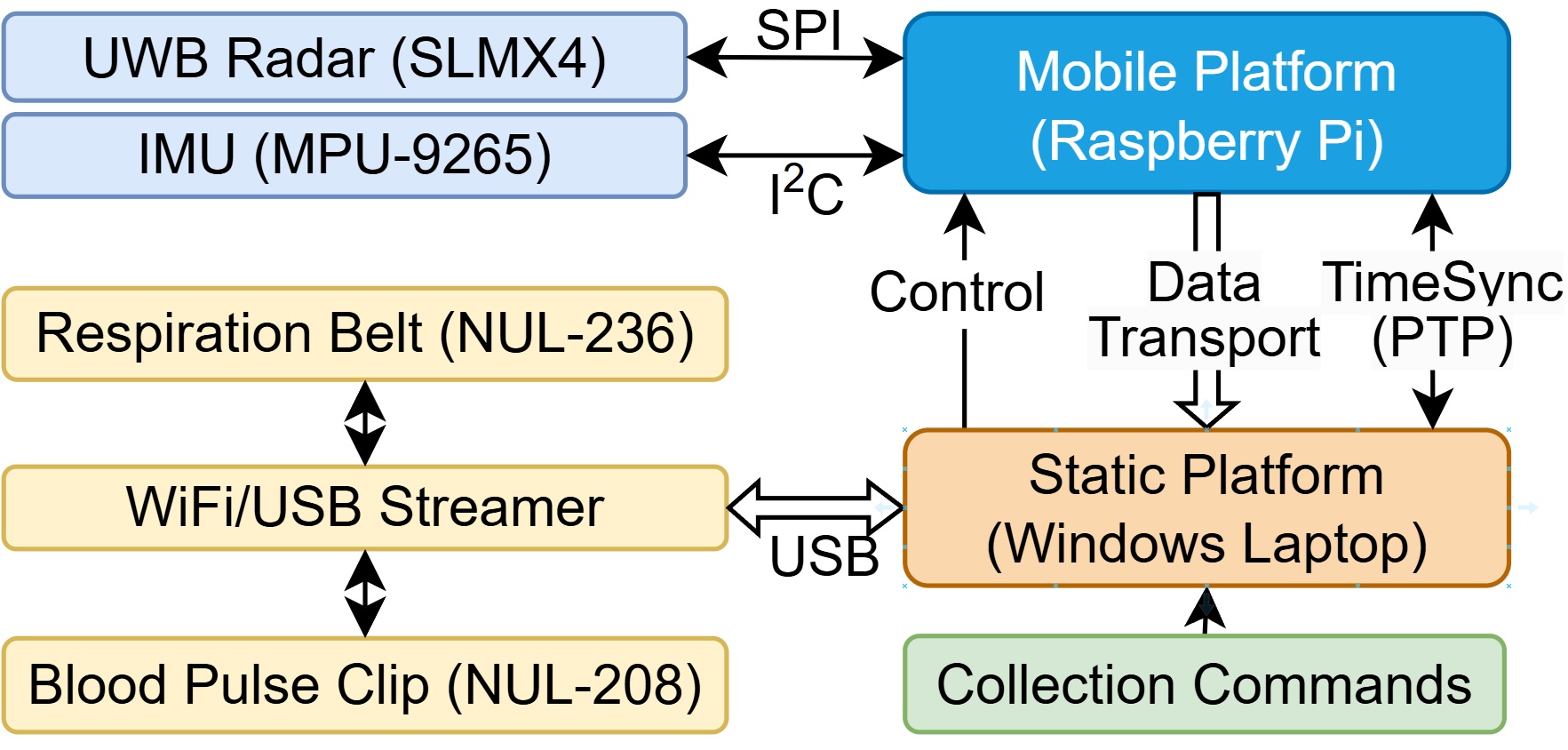}
    \vspace{-8pt}
    \caption{\name data collection platform.}
    \vspace{-10pt}
    \label{fig:platform}
\end{figure}

As a dataset focusing on measuring human vital signals (emphasizing respirations), our dataset collection platform is centered around an IR-UWB radar, as is shown in Figure~\ref{fig:platform}. The UWB radar of our choice is the SLMX4~\cite{SLMX4Nov24:online} manufactured by SensorLogic. The IR-UWB radar combines low power consumption with high resolution and thus has great potential to be integrated into mobile platforms. Instead of transmitting continuous waves, the UWB radar unit transmits ultra-short Gaussian pulses centered at 7.29~\!GHz with a bandwidth of 1.4~\!GHz. The device is FCC-approved to operate with a transmission power of -41.3~\!dBm/MHz, making the total transmission power less than 0.1~\!W. The 1.4~\!GHz wide bandwidth also provides a relatively high resolution of $\frac{c}{2B} = \frac{3e^9 m/s}{2\times 1.4e^9 Hz} = 10.7 cm$, allowing for separating respiration signals from multiple subjects. 

\begin{figure}
    \centering
    \includegraphics[width=0.85\linewidth]{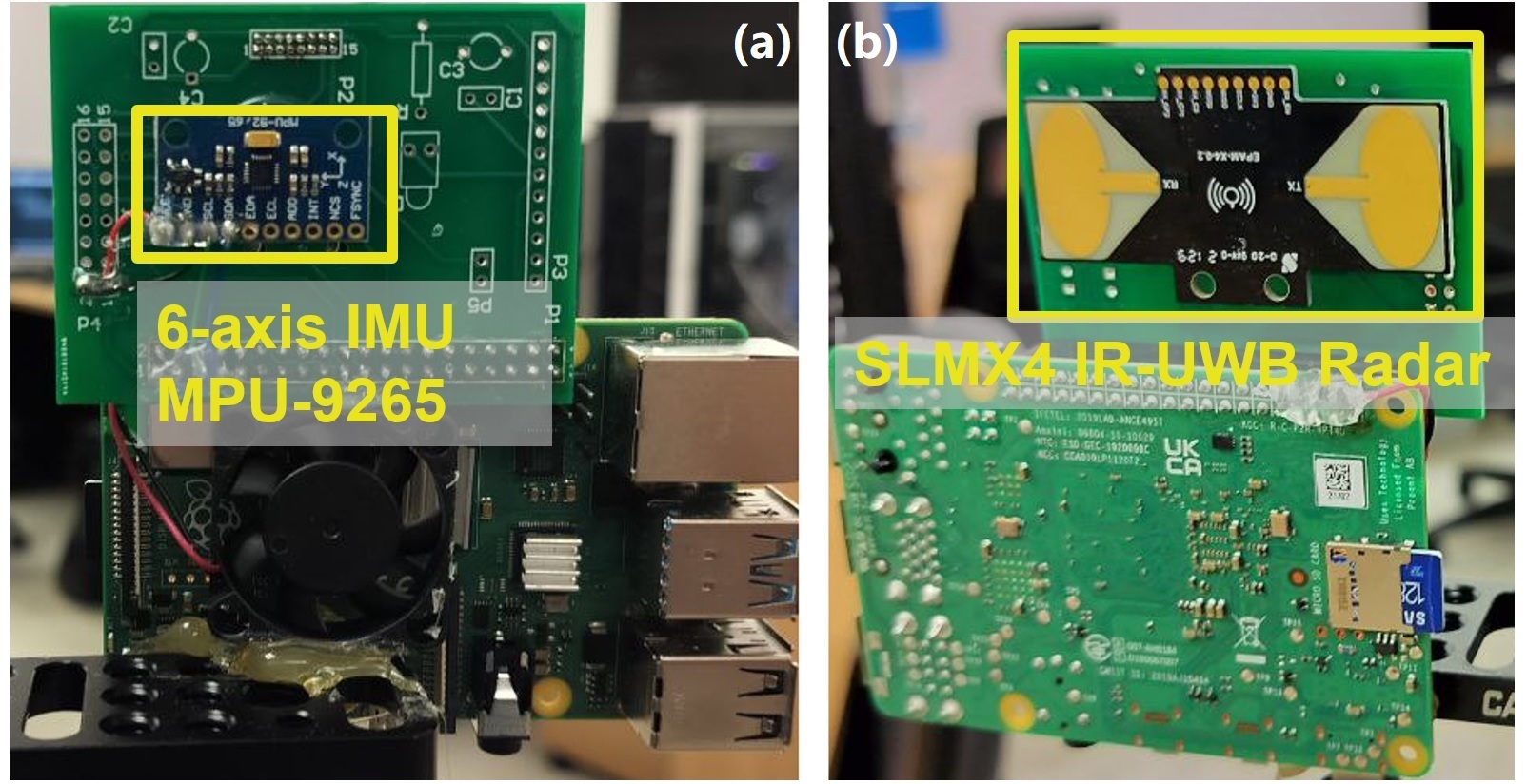}
    \vspace{-10pt}
    \caption{Mobile platform hardware of \name: (a) back-side. (b) front-side.}
    \vspace{-15pt}
    \label{fig:hardware}
\end{figure}

In our experiments, we configure the radar to repeat the pulses at 50~\!Hz, and focusing on the range between 0.3 to 6.3 meters. The SLMX4 radar sensor of our choice has a spatial precision of 5.14~\!cm, which means the 6~\!m space is divided into 120 different distance bins. The resulting data is $D_{uwb} \in \mathbb{C}^{120 \times 50t}$, where $t$ is the time in seconds, and each datapoint is a complex number. The UWB radar is connected to and controlled by a Raspberry Pi using a SPI interface. An inertial measurement unit (IMU) is also mounted on the Raspberry Pi using the I2C interface to record the movement sensor platform unit at a sample rate of 100~\!Hz. The resulting data is $D_{imu} \in \mathbb{R}^{6 \times 100t}$, where 3 axes are used for the accelerometer to record linear movements and the remaining 3 axes are used by the gyroscope to record rotations. The Raspberry Pi 4B, the IR-UWB radar, and the IMU form the mobile platform. They are mounted together on a cheeseplate (see Figure~\ref{fig:hardware}), which can be either placed on a tripod or held in hand. 

To collect ground truth respiration waveform, we leverage a wearable sensor, NeuLog respiration monitor belt logger sensor NUL-236~\cite{Respirat93:online}, which is widely adopted by many related work as a ground truth sensor~\cite{ren2015noncontact, kam2017all, rahman2021estimation, zheng2020v2ifi}.
The sensor is a belt mounted on the subject's chest, with some air pumped inside. The air pressure inside the belt will change according to the subject's chest movement. The air pressure is converted to electronic signals by a piece of silicon between two layers of metal foil whose resistance will change following the pressure on it. Although not a primary focus of this paper, our dataset employs a heart rate and pulse logger sensor NUL-208, which collects the ground truth blood pulse waveform using a finger clip containing plethysmograph-based electrodes. This ground truth is helpful to those works seeking to reconstruct heart rate or blood pulses. The two NeuLog sensors both operate at 50~\!Hz, generating $D_{heart} \in \mathbb{R}^{1 \times 50t}$ and $D_{breath} \in \mathbb{R}^{1 \times 50t}$. The ground truth sensors are connected to a NeuLog USB steaming unit, controlled by an HTTP-based API on a Windows laptop. The Neulog sensors and the Windows Laptop form our system's static platform. 

The Windows laptop and the Raspberry Pi are time synchronized using Precision Time Protocol (PTP), with the Raspberry Pi being the server. During data collection sessions, control scripts on the Windows laptop start all the static and mobile sensors. Each session lasts around five minutes, which are further divided into 30-second sub-sessions, constrained by the buffer size of the ground truth sensors. Each sensor's data are timestamped by the local machine and transported to the Laptop at the end of each session. All the data are resampled to 50~\!Hz and aligned based on timestamp proximity (using the UWB timestamp as a reference).

\subsection{Data Collection Protocol and Statistics}
Now, we provide details about our data collection protocol. Our study is IRB-approved (UCLA IRB\#23-000754). We recruited 12 participants for our study. Each experiment visit consists of four five-minute sessions, and participants may choose to attend an arbitrary number of visits. During the data collection, the participants were instructed to sit on a chair 1.5 meters away from the sensor and breathe normally. Participants are advised to relax and avoid controlling their breath or making body movements for about 20 minutes while data is collected. They may choose to watch videos, listen to music, or meditate. Next, a researcher assisted in fitting wearable sensors: a heart rate and pulse sensor clipped to the left-hand pinky finger, and a respiration monitor belt secured around the lower ribs and diaphragm, inflated to a comfortable fit. 

\begin{figure}[h!]
    \centering
    \includegraphics[width=\linewidth]{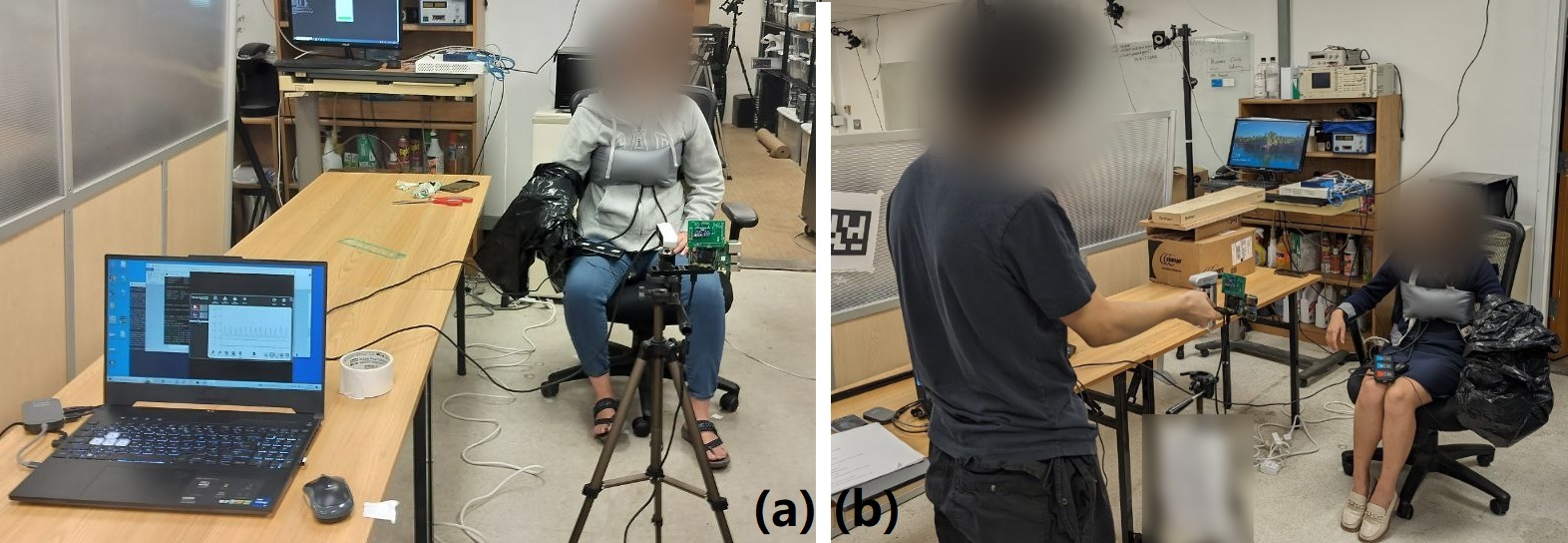}
    \caption{Experiment scenarios: (a) tripod, (b) handheld.}
    \label{fig:exp}
\end{figure}

The experiment platform was placed on a tripod with the UWB radar facing the subjects' chest, as depicted in Figure~\ref{fig:exp}(a). Also, some recent research has emerged to focus on the challenges caused by the relative motion between the sensor and the subject~\cite{zhang2022mobi2sense,chang2024msense}, we also include a handheld case where a research is holding the sensing platform in hand, with involuntary hand motion blended into the data (see Figure~\ref{fig:exp}(b). While not a primary focus of this paper, we will release this part of the data together with the tripod case. The dataset statistics is shown in Table~\ref{tab:stats}. In total, we collected 937 minutes of tripod data from 12 subjects and 546 minutes of handheld data from 11 subjects, which ensures both diversity across users and diversity across different respiration patterns on different days. This 24-hour dataset is fully anonymized and will be released to benefit the research community.

To verify the quality of the dataset, we also performed a simple benchmarking test using the DeepVS~\cite{xie2022deepvs} model, which is a transformer-based model that estimates the breath rate and heart rate simultaneously in an end-to-end manner. The DeepVS model first extracts time and frequency domain features from the signal using 1D-CNN layers and then applies an attention module to combine these features. Two output layers are used to generate heart rate (HR) and respiration rate (RR) estimations. Our results show about a 2 beats-per-minute (bpm) error for RR, and a 6.7 bpm error for HR, which is close to that reported in the DeepVS paper~\cite{xie2022deepvs}. Note that in the handheld case, we see a much larger RR error (this is expected due to signal quality degradation caused by the hand jitter), but a slightly lower HR error. We presume this is because we directly use the hyper-parameters specified in the paper. More specifically, a $\lambda$ hyper-parameter is applied to balance the loss calculated from the HR branch and the RR branch, and it may take some additional effort to find the optimal $\lambda$. Nevertheless, this benchmarking experiment demonstrates that our data contains information related to HR and RR and can be used to develop related models and algorithms.

\begin{table}[h!]
\begin{tabular}{|c|c|c|c|c|}
\hline
         & \begin{tabular}[c]{@{}c@{}}Mean RR\\ Error (bpm)\end{tabular} & Std  & \begin{tabular}[c]{@{}c@{}}Mean HR\\ Error (bpm)\end{tabular} & Std  \\ \hline
Tripod   & 1.99                                                          & 2.05 & 6.67                                                          & 5.46 \\ \hline
Handheld & 3.39                                                          & 2.76 & 5.81                                                          & 4.98 \\ \hline
\end{tabular}
\caption{Dataset benchmarking results using the DeepVS~\cite{xie2022deepvs} model.}
\vspace{-15pt}
\label{tab:bench}
\end{table}

%% file: Contents/04-Method.tex
\section{\textbf{System Design}}\label{Sec:design}
In this section, we introduce the design of \name in detail. We will introduce the system overview, the breathing waveform inversion detector algorithm, and the architecture of the autoregressive signal quality predictor along with their design intuitions.

\subsection{Overview}
\begin{figure}[h!]
    \vspace{-5pt}
    \centering
    \includegraphics[width=0.7\linewidth]{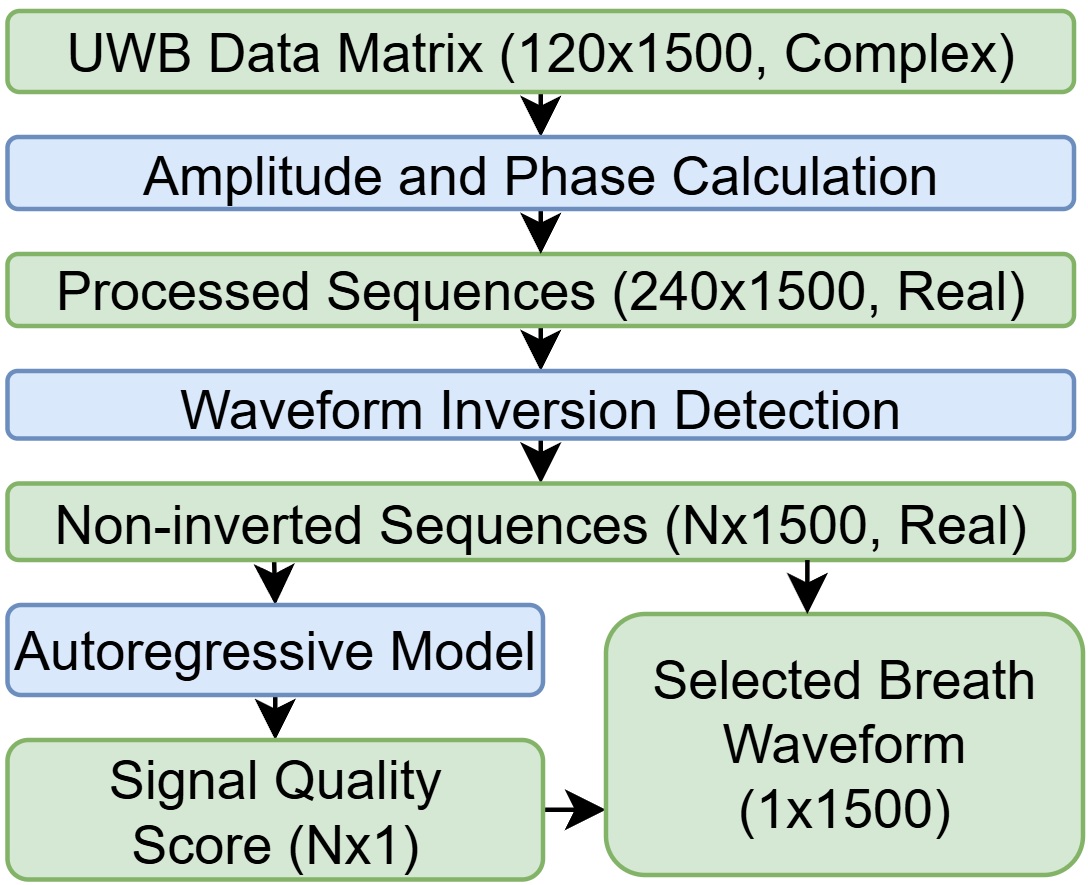}
    \vspace{-5pt}
    \caption{Overview of the \name system design.}
    \label{fig:system-design}
    \vspace{-10pt}
\end{figure}

The system design for \name is illustrated in Figure~\ref{fig:system-design}. The input to the system is the entire complex UWB matrix (120 x 1500). We have shown that both the magnitude and phase of a UWB signal contain respiration waveform information in Section~\ref{sec:prelim}. Thus, \name first computes the magnitude and the phase along each distance bin, resulting in 240 possible candidates. The phase is calculated based on the angle of the complex I/Q value, plus an unwrap operation to mitigate gaps around $2\pi$. The 240 candidates are then filtered using our Inversion Detector, Algorithm~\ref{alg:inv}, removing any sequences classified as inverted. Each remaining candidate sequence is then scored by the autoregressive predictor. Finally, \name selects the sequence with the highest MobiVital score from the remaining candidates, returning the selected sequence and the score.

\subsection{Inversion Detector}

\begin{algorithm} [b]
    \footnotesize
    \SetKwComment{tcp}{//}{}
    \caption{\small{Breath Waveform Inversion Detector}}
    \label{alg:inv}
    \KwIn{Candidate time series $ y \in \mathbb{R}^{T}, r_{th}=1.0 $}
    \KwOut{Indicator of Waveform Inversion $inv \in [0,1]$}

    $y_{s} \leftarrow SavitzkyGolayFilter(y, order=5, frame\_len=100)$ \tcc*{Smooth the original waveform} \par
    $\boldsymbol{P_{pos}}  \leftarrow FindPeaks(y_s, prominence=0.1)$ \par
    $\boldsymbol{W_{pos}} \leftarrow PeakWidth(y_s, \boldsymbol{P_{pos}}, rel\_height=0.5)$ \par
    $w_{pos} \leftarrow Average(W_{pos}) $ \par
    $\boldsymbol{P_{inv}}  \leftarrow FindPeaks(-y_s, prominence=0.1)$ \par
    $\boldsymbol{W_{inv}} \leftarrow PeakWidth(-y_s, \boldsymbol{P_{inv}}, rel\_height=0.5)$ \par
    $w_{inv} \leftarrow Average(W_{inv}) $ \par
    \algorithmicif $w_{pos} / w_{inv} < r_{th}$ \algorithmicthen{\text{   }$inv \leftarrow 0$  } \algorithmicelse{\text{   }$inv \leftarrow 1$}
    \end{algorithm}
    
The UWB candidate time series may be inverted with respect to the chest movement direction. The problem is overlooked by previous works as their primary focus is the breathing rate. With our goal being to produce high-quality waveforms, the inversion problem must be addressed. To this end, we propose a biology-informed algorithm to detect inverted UWB sequences, leveraging the morphology of respiration. The inspiration for the algorithm comes from the biological nature of human respiration. During inhalation, chest, and abdominal muscles contract to draw air into the lungs. Upon exhalation, these muscles relax, allowing the lungs to deflate~\cite{national2022how}. Since holding the air in the chest requires active muscle effort, normal respiration waveform often shows spikes for active inhalation-exhalation, with longer gaps between the spikes. In other words, if we denote the deflated chest location to be ``0'' and the inflated chest location to be ``1'', then the ``duty cycle'' $\frac{t_1}{t_1 + t_0}$ should be less than 0.5.

Inspired by this fact, we proposed the algorithm detailed in \ref{alg:inv}. In this algorithm, we first use a Savitzky-Golay finite impulse response (FIR) smoothing filter with a polynomial order of 5 and a frame length of 100 to process the candidate sequence to remove its jitters and microspikes. After smoothing, we find the peaks on the smoothed signal and calculate its average peak width $W_{pos}$. Here the peak width is calculated at height $h = h_{peak} - 0.5P$, where P is the prominence (measures how much a peak stands out from its surroundings) of the peak~\cite{peakwidt69:online}.
The same procedure is applied to the flipped version of the signal $W_{inv}$. We determine the signal to be inverted if $W_{pos}/  W_{inv} \leq r_{th}$, or not inverted otherwise.

\begin{figure}[t!]
    \vspace{-15pt}
    \centering
    \includegraphics[width=\linewidth]{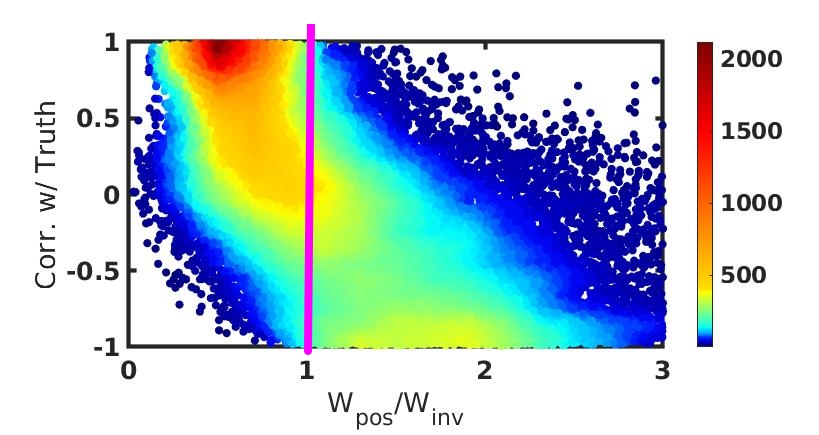}
    \vspace{-20pt}
    \caption{Peak width statistics of the sequences in the training dataset.}
    \label{fig:w2w1}
    \vspace{-10pt}
\end{figure}

\begin{figure}[t!]
    \centering
    \includegraphics[width=\linewidth]{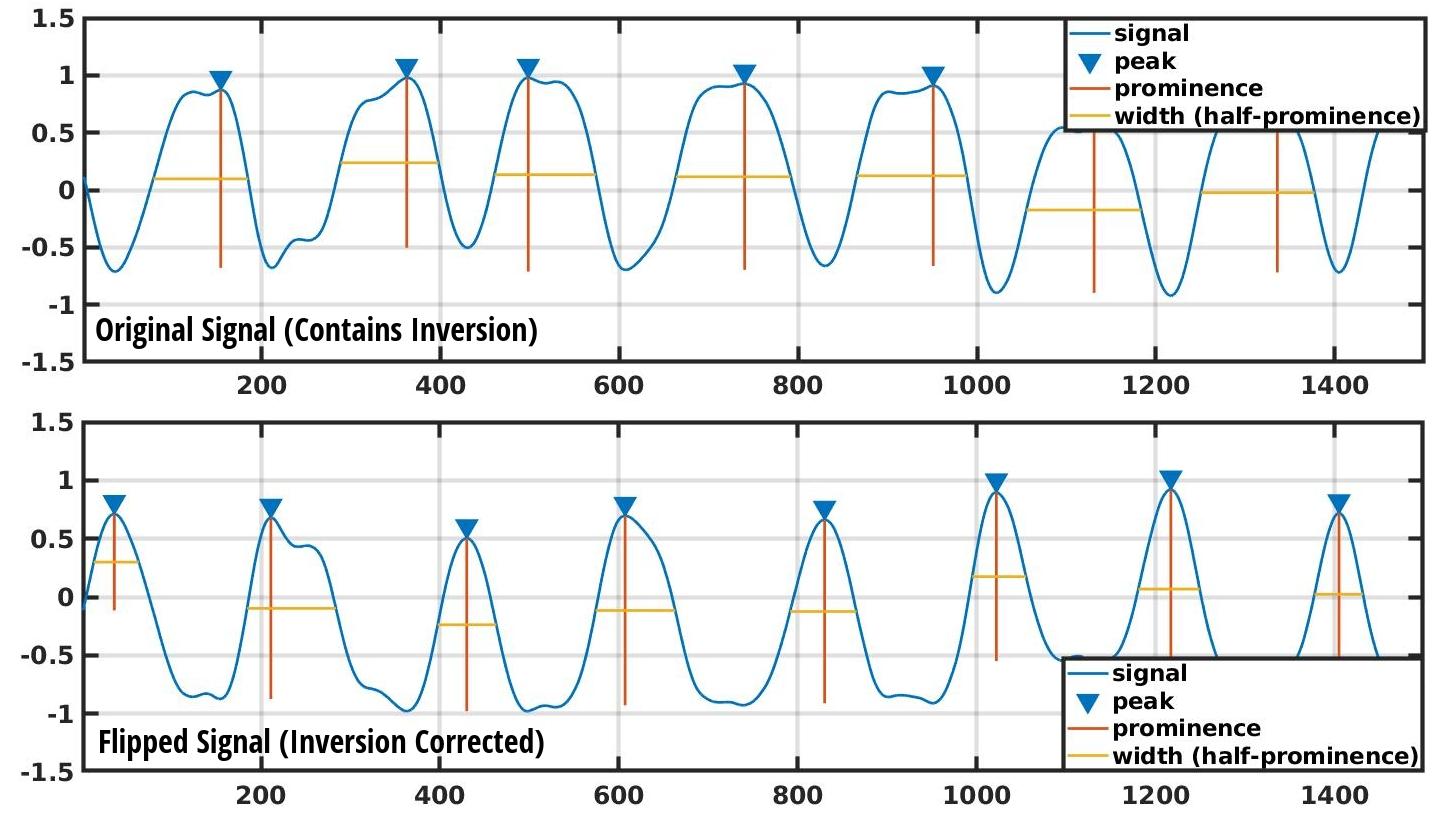}
    \vspace{-15pt}
    \caption{Example of the breath waveform inversion detector. The upper original signal has an average peak width of 119.46 and the mirrored version below has 70.17. The signal is considered to be inverted.}
    \label{fig:invert-func}
    \vspace{-15pt}
\end{figure}

\begin{figure*}
    \centering
    \includegraphics[width=\linewidth]{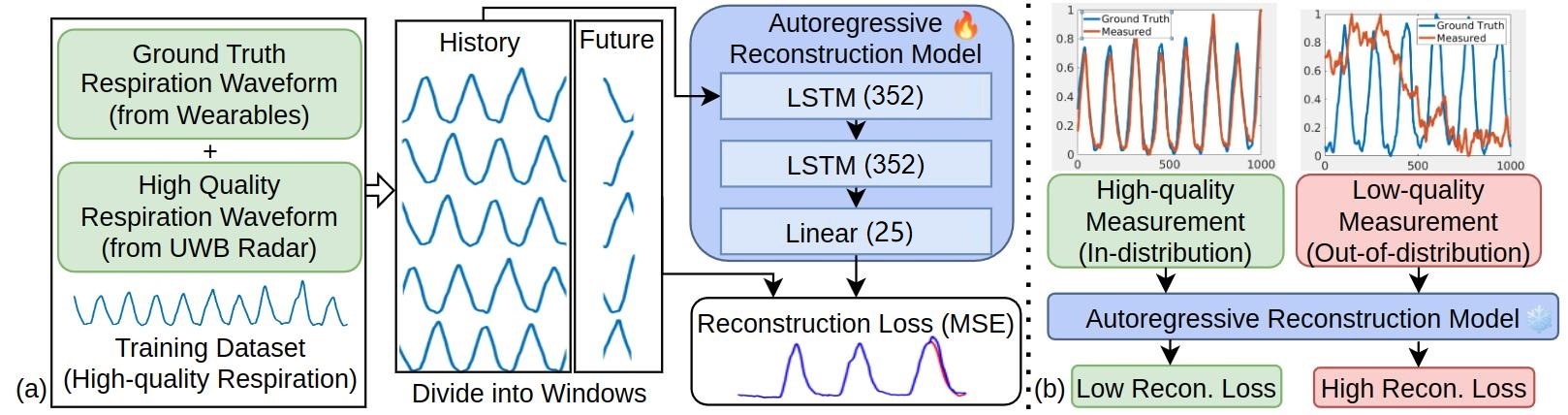}
    \vspace{-15pt}
    \caption{Autoregressinve predictor. Part A shows the autoregressive model training pipeline. Part B shows the model's behavior during deployment when exposed to high and low-quality input sequence examples.}
    \label{fig:autoregressive}
     \vspace{-10pt}
\end{figure*}

We statistically prove the design intuition of our inversion detection algorithm in Figure~\ref{fig:w2w1}. We calculate the ratio of $\frac{W_{pos}}{W_{inv}}$ and plot it against the signal quality measurement (correction coefficient with the ground truth). We can see that $r_{th} = 1$ draws a line that separates the red cluster consisting of high-quality signals and the bottom green cluster composed of inverted signals. Furthermore, we provide a numerical example of Alg.~\ref{alg:inv} in Figure~\ref{fig:invert-func}. The upper side of this example figure shows the original signal, with peaks indicated by blue triangles and peak width indicated by yellow lines. The lower side shows the same for the flipped signal. We can see that $\frac{W_{pos}}{W_{inv}}=\frac{119.46}{70.17}>1$, thus the original signal is considered inverted.\looseness=-1

There are two ways we can use our inversion algorithm which we call pre-inversion and post-inversion. For pre-inversion, we first filter the UWB and remove the candidacy of any inverted sequences. For post-inversion, we first use some algorithm to pick the best time series and flip the sequence if it's considered inverted. For \name, we determined that pre-inversion is a better choice as the cost of wrongly flipping a non-inverted signal is high.


\subsection{Autoregressive Predictor}

MobiVital leverages an autoregressive deep learning model to automatically select the best sequence to represent the users respiration, among all the candidate time series. The intuition of this method is leveraging the tendency of deep neural networks to overfit the distribution of their training datasets. This tendency is often considered to be detrimental, causing neural networks to work well on data with a similar distribution to the training dataset, and work poorly if the distribution of data is different. However, we see opportunities in this generalization limitation. If we can train a neural network that learns the dynamics or the distribution of high-quality respiration data, the neural network's performance becomes an indicator: a good performance suggests ``in distribution'' (high signal quality).

The autoregressive model \textit{training} pipeline is detailed in part a of Figure~\ref{fig:autoregressive}(a). The word ``autoregressive'' means the task of the model is to reconstruct part of the sequence given some history values. Since we want the model to learn the dynamics of respiration waveforms, we construct a training dataset using (1) ground truth respiration waveform from the on-body sensors and (2) high-quality UWB-measured respiration waveform. High-quality sequences are defined as sequences that have a correlation coefficient with the ground truth respiration waveform higher than a threshold $r_0$. We pick $r_0 = 0.9$. During the development of \name, we also swept $r_0$ from 0.5 to 0.95, and the performance remained similar.

The sequences in the training dataset are then processed using a sliding window of 4.5 seconds. Inside the window, the first 4 seconds (200 samples) become the ``history'', and the remaining 0.5 seconds (50 samples) become the ``future''. Then the sliding window moves forward in increments of 0.5 seconds to generate more histories and futures. Our autoregressive reconstruction model is a lightweight 2-layer LSTM model followed by a linear layer. The model takes the history as the input and predicts the future sequence of 0.5~\!s. The Mean Squared Error (MSE) between the predicted segment and the future segment is used as the loss function. This MSE loss is only used during training.

Figure~\ref{fig:autoregressive}(b) shows the autoregressive model's behavior during deployment time (with the model's weights frozen). The model accurately reconstructs the time series if the candidate measurement is high-quality, i.e., it ``looks like'' the respiration signals the model was trained on. However, if the input sequence is low-quality and the model was not trained on similar data, the model struggles to reconstruct the sequence, resulting in a high reconstruction loss.


\subsection{MobiVital Score}
As the last step, the \name score is a metric produced by the autoregressive model used to determine the quality of each possible candidate sequence. Algorithm~\ref{alg:score} describes the calculation algorithm. The signal $y$ is firstly chopped into a history set $T^{his}$ and a future set $T^{fut}$. For all the entries in $T^{his}$, we use the autoregressive model to predict a future $t_{pred}$. Then, we calculate the correlation coefficient $r_{c,i}$ between $t_{pred}$ and $T_{i}^{fut}$. Averaging over all the correlation coefficient $r_{c,i}$'s gives us the \name score $m_s \in [-1,1]$. In this algorithm, no additional ground truth signal is used and the signal waveform itself serves as the ground truth, that is why we call the pipeline ``self-supervised''.

\begin{algorithm}[t!]
    \footnotesize
    \SetKwComment{tcp}{//}{}
    \caption{\small{\name Score Calculation}}
    \label{alg:score}
    \KwIn{Candidate time series $ y \in \mathbb{R}^{T}$}
    \KwOut{Mobivital score $m_s \in [-1,1]$}

    $T^{his}, T^{fut}\leftarrow SlidingWindow(y)$ \par
    $m_s \leftarrow 0 $, $len \leftarrow ||T^{his}||$ \par
    LOOP FOR $i$ FROM $1$ TO $len$ \par
        \text{ } \text{ } $t_{pred} \leftarrow AutoregssiveModel(T^{his}_i)$ \par
        \text{ } \text{ } $m_s \leftarrow m_s +  CorrCoeff(T^{fut}_i, t_{pred}) / len $ \par
    \end{algorithm}

\begin{figure}[t!]
    \vspace{-15pt}
    \centering
    \includegraphics[width=\linewidth]{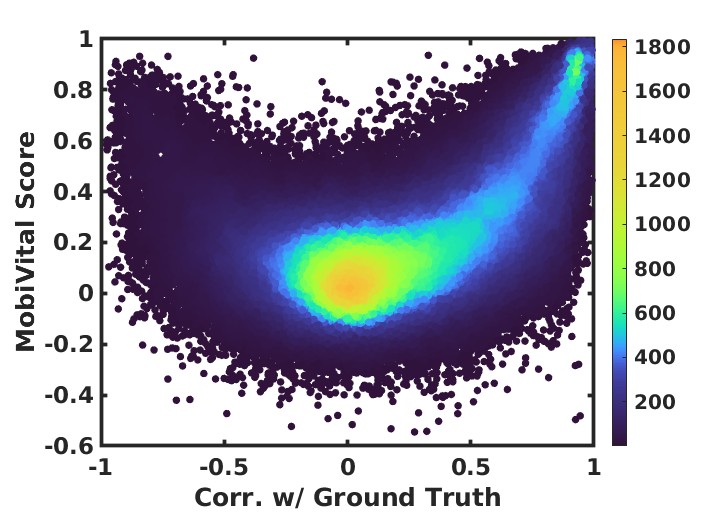}
    \vspace{-15pt}
    \caption{MobiVital score on all candidate time series. This figure shows that \name score can be a good surrogate of the time series quality.}
    \label{fig:mobivital-rawscore}
     \vspace{-5pt}
\end{figure}

For the score to be a good surrogate of the signal quality, it must ideally be bounded and have a positive linear relationship with the ground truth correlation. The \name score satisfies both of these criteria. Figure~\ref{fig:mobivital-rawscore} shows the \name score on the training users having an almost 1 to 1 relationship with the ground truth correlation. The score is also strictly bounded from -1 to 1 inclusive. Thus, the \name score can be used to measure the quality of the signal, without any knowledge of the ground truth respiration waveform. 

%% file: Contents/05-Results.tex
\section{\textbf{Evaluations}}\label{Sec:results}

\subsection{\textbf{Implementations}}
\subsubsection{Dataset Creation}
Our dataset contains 12 subjects, with around 1 hour of data per subject when the sensor is tripod-mounted. To thoroughly test the generalization ability methods, we spare the data from users G, H, I, and J for the final evaluations. The remaining 8 users' data form the train-dev dataset that is used to train the autoregressive model and to optimize the hyper-parameters. 


\subsubsection{Hyperparameter Tuning}
The training and model hyperparameters are tuned using Mango~\cite{sandha2020mango}, a parallel hyperparameter tuning library leveraging Bayesian optimization. This significantly reduces the search time over the large hyperparameter space. The optimized hyperparameters include batch size, epochs, learning rate, history length, future length, LSTM hidden size, and the number of LSTM layers.

The hyperparameter sets were evaluated as follows. Each set of parameters trains a model on $6/8$ subjects of the training set and is evaluated on the remaining 2 subjects. The objective function is the average correlation between the model's selected sequences and the ground truth, averaged across three random seeds to account for the initialization variability of LSTMs. The final selected hyperparameters by Mango are in Table \ref{tab:hyperparameters}.

\begin{table}[h!]
    \vspace{-10pt}
    \centering
    \begin{tabular}{l c}
        \toprule
        Parameter & Value \\
        \midrule
        Batch size & 64 \\
        Epochs & 20 \\
        Learning rate (lr) & 1 * 10\textsuperscript{-4} \\
        History sequence length & 200 (4 seconds) \\
        Future sequence length & 25 (0.5 seconds) \\
        LSTM hidden size & 352 \\
        Number of LSTM layers & 2 \\
        \bottomrule
    \end{tabular}
    \caption{Tuned Hyperparameters by Mango}
    \vspace{-30pt}
    \label{tab:hyperparameters}
\end{table}

\subsection{Metric and Baselines}

\subsubsection{Metric}

Correlation coefficients have been mentioned throughout this paper as the key metric for measuring the quality of a UWB sequence compared to the ground truth. Given measurement $x[n]$ and ground truth $y[n]$, the correlation coefficient $r$ is calculated as
\begin{equation}
    r = \frac{\sum (x_i - \bar{x})(y_i - \bar{y})}{\sqrt{\sum (x_i - \bar{x})^2} \cdot \sqrt{\sum (y_i - \bar{y})^2}}.
\end{equation}
The correlation coefficient describes if the relationship between $x[n]$ and $y[n]$ can be described using a linear equation. i.e., $r=1$ when $x[n] = ay[n]+b$. Intuitively, $r$ describes if $x$ and $y$ are following the same trend, e.g., if the chest movement $y$ shows inhalation, whether $x$ follows the same pattern. The correlation coefficient is sensitive to time-shift, noise, and distortion, but is robust to constant value offset or scaling. This characteristic is helpful for us, as we want the UWB waveform to have the same trend as the wearable sensor, but the absolute values do not matter as the signals are normalized. The correlation coefficient is also bounded in [-1,1], making it ideal to score signal quality compared with L1-norm or L2-norm.

\subsubsection{Baselines}
As we have introduced in Section~\ref{Sec:intro}, existing UWB respiration monitoring systems have proposed different methods to slice the UWB data and get the best sequences representing the respirations. We pick three methods dominantly used in literature as baselines, namely Variance \cite{yang2019multi, kim2019non,husaini2022non}, Signal to Noise Ratio (SNR) Estimation~\cite{liu2022vital}, and Constant False Alarm Rate (CFAR)~\cite{zheng2020v2ifi, chang2024msense}. These signal processing algorithms share the same pre-processing along the time axis: a loop-back filter is first applied to remove clutters from static objects~\cite{zheng2020v2ifi}. Then a detrend algorithm is applied to remove the polynomial trend in the signal. The three baseline methods diverge afterward. Below are the key details of these baseline methods.

\begin{itemize}
    \item Variance: For each distance bin $d$, the algorithm slice out the correspond signal $x_d(t)$. Then, the variance $Var(x_d(t))$ are calculated. Finally, a peak detection algorithm is performed on the variance vector $\mathbf{V} \in \mathrm{R}^D$, and the bin with the highest peak is selected as the target bin. This method only searches on the amplitude of the UWB matrix.
    \item CFAR: For each distance bin $d$, the algorithm performs FFT on the time series $x_d(t)$, resulting in an range-FFT map. Afterward, a Constant False Alarm Rate (CFAR) algorithm is used to search for the most significant peaks on the map. The distance bin contains the pick that was selected. This method also searches on the amplitude of the UWB matrix only.
    \item SNR: An FFT is performed on the time series $x_d(t)$ similar to CFAR. By calculating the ratio of energy in the frequency band corresponding to respiration (0.2-0.7~\!Hz) to the energy in the remaining frequency band, one can estimate the signal-to-noise ratio. The bin with the highest SNR is selected. This method can search on both the amplitude and the phase of the UWB matrix.
\end{itemize}

For fairness, our inversion detected algorithm \ref{alg:inv} is applied to the baseline methods. CFAR and Variance both use post-inversion as their peak detection algorithms require peak detection on the entire landscape so we cannot remove these assumed inverted signals before executing the algorithm. \name and SNR use pre-inversion since they can individually score each sequence.

Finally, an oracle demonstrates the best possible score given knowledge of the respiration waveform from the wearable band. This is not achievable in practice since it requires knowledge of the ground truth but is included to show the upper theoretical bound. The oracle selects the best choice for each instance by choosing the sequence with the highest correlation to the ground truth.


\subsection{\textbf{Quantitative Results}}
Next, we move on to see some numerical results. We first examine the quality of the time series selected by \name.
\subsubsection{Quality of the signals selected by \name.}
\begin{figure}[h!]
    \vspace{-5pt}
    \centering
    \includegraphics[width=0.9\linewidth]{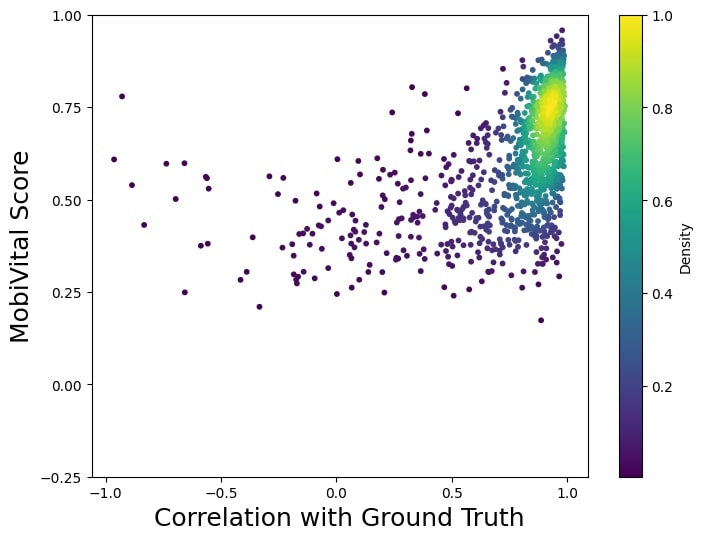}
    \vspace{-5pt}
    \caption{MobiVital score on sequences selected in the test set. This figure shows that the MobiVital score can measure quality, even on unseen users.}
    \label{fig:mobivital-score}
    \vspace{-10pt}
\end{figure}

In Figure~\ref{fig:mobivital-rawscore}, we have shown the \name score on all the candidate time series. Our system selects the time series with the highest \name score. Those points corresponding to the selected time series are kept in Figure~\ref{fig:mobivital-score}. We can see that most of the selected time series have a high correlation with the ground truth, with an average correlation coefficient of 0.819. The points are concentrated in the top-right corner of the figure, where both the signal quality and the score are high. These results show that \emph{the \name score can be used as a good surrogate for the ground truth correlation to pick high-quality signals.} Another advantage of \name is that it generates an absolute score bounded by [-1,1]. While the SNR method can also produce a metric, its score is not bounded, ranging from $-\infty$ to $\infty$. The other two baseline methods are relative metrics. Having an absolute, bounded metric is important as it can inform the downstream modules when the signal is of low quality and no good respiration measurement is available.


Figure \ref{fig:cdf} shows the CDF of each method's performance. There is a clear separation between the baselines and \name. Our improvement stems from selecting fewer sequences with inversion. Less than 5\% of the sequences selected by \name have a negative correlation, compared to 10\% for the closest baseline method. Compared to the more popular methods of CFAR and Variance, the improvement achieved by \name becomes even more evident.

\subsubsection{Comparison with the baselines.} 
\begin{figure}[h!]
\vspace{-15pt}
    \centering\includegraphics[width=0.9\linewidth]{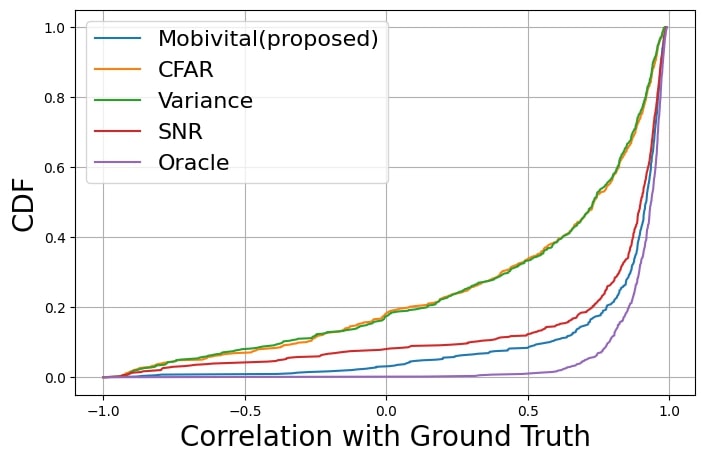}
    \vspace{-10pt}
    \caption{CDF of methods with the inversion detection algorithm.}
    \label{fig:cdf}
\end{figure}

In Table~\ref{tab:results} and Figure~\ref{fig:cdf}, we compare the performance of \name with the three baselines. Figure~\ref{fig:cdf} shows a cumulative distribution function (CDF) on the quality of the signals selected by each method, and Table~\ref{tab:results} compliments the figure with the mean correlation coefficient obtained by each method. From the CDF curve, we can see that \emph{\name has a clear advantage over the baselines} in terms of the waveform quality and is the closest to the oracle. The statistics in Table~\ref{tab:results} suggest that \name improves from the closest baseline, SNR, by 7\%, achieving a score of 0.819. Compared to SNR without pre-inversion, the gap increases to 34\%. Note that the inversion detector is also a contribution of this paper. The advantage is even larger for the other two baselines. A possible reason is that both Variance and CFAR are designed to search for the bin containing the strongest human body reflections. However, such a bin is often suboptimal, leading to the inferior but similar performance of these two baseline methods.




\begin{table}[h!]
\centering
\begin{tabular}{lcc}
\toprule
\textbf{Method} & \textbf{w/ Inv Det.} & \textbf{w/o Inv Det.} \\ 
\midrule
\name & \textbf{0.819} & \textbf{0.816} \\
SNR       & 0.745 & 0.475 \\
CFAR      & 0.516 & 0.218 \\
Variance  & 0.514 & 0.225 \\
Oracle    & - & 0.943 \\ 
\bottomrule
\end{tabular}
\caption{Average score of methods}
\label{tab:results}
\end{table}
 
\subsubsection{Ablation study on the inversion detector.} 
\begin{figure}[t!]
    \centering
    \includegraphics[width=1\linewidth]{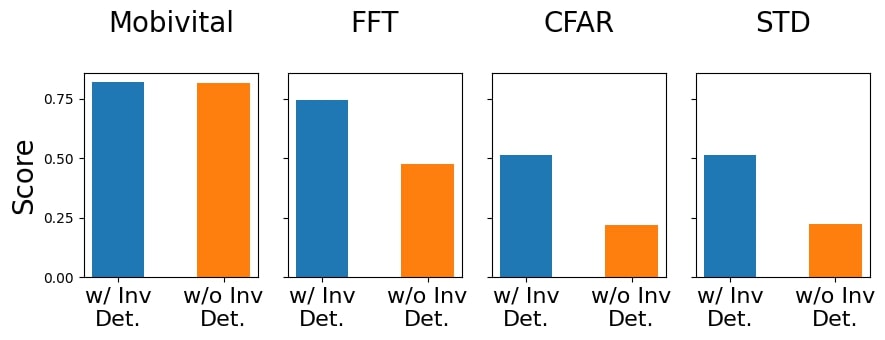}
    \caption{Comparison of methods with and without the inversion detection algorithm.}
    \label{fig:inversion}
\end{figure}

Next, we use ablation studies to demonstrate the effectiveness of our signal inversion detection algorithm.
Table~\ref{tab:results} and Figure~\ref{fig:inversion} show the improvement the inversion algorithm brings. \emph{We can see a clear performance gap if the algorithm is removed}. The inversion algorithm doubles the performance for CFAR, Variance, and SNR. MobiVital's improvement is minor, with only a 0.03 increase. This is because we trained the \name model with high-quality signals without any inversion. Thus, the inverted signal is also out-of-distribution for the autoregressive model in \name, enabling the system to naturally reject some of the inverted candidates even without the inversion algorithm. Nevertheless, performing pre-inversion on \name is still beneficial for performance by reducing the possibility of picking inverted signal candidates. The inversion detection algorithm is very lightweight with minimal computation overhead.\looseness=-1



\subsubsection{Inter-user variability.}
\begin{figure}[b!]
    \centering
    \includegraphics[width=0.9\linewidth]{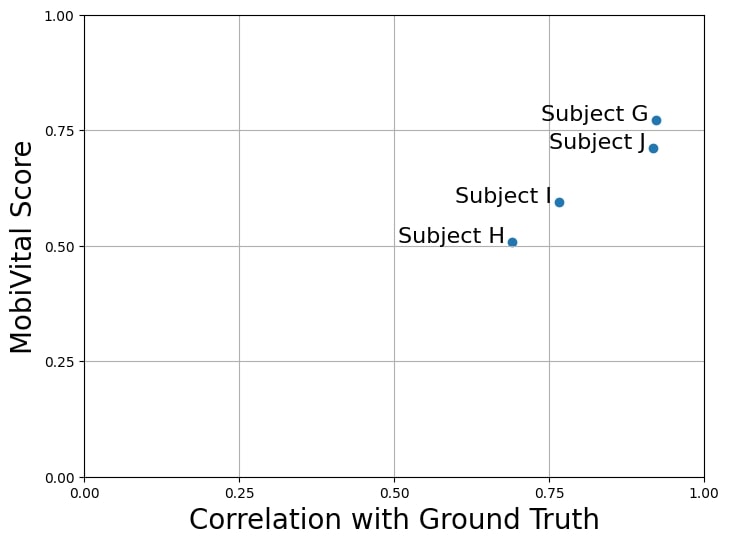}
    \caption{Scores of individual unseen test subjects.}
    \label{fig:test-users}
\end{figure}

We also individually analyzed the performance of \name on the test subjects: G, H, I, and J to study subject variability. Figure \ref{fig:test-users} shows the average correlation with the ground truth plotted against the average \name score for each subject. First, the quality of the time series selected by \name is greater or equal to 0.7. Second, among the four users, we found users G and J's data exhibits a higher quality, while users I and H's data quality is worse. As a consequence, the \name score of I and H is also lower. These results reiterate that \name score can represent the signal quality effectively. 

Many possible reasons could contribute to the inter-user variability. One possibility is involuntary body movement. While we instruct the users to keep their body pose unchanged during sessions, involuntary body movements such as nodding, kicking, or adjusting sitting positions may happen and contaminate the respiration signal. While tackling the body motion is beyond the scope of this work, some pioneer works have proposed different solutions~\cite{zheng2021more, chen2021movi}. Second, the specific breathing patterns of that subject may be out of the distribution of the training dataset. Each person has a unique breathing habit that produces slightly different patterns. Our training set contains 8 users, and the breathing patterns of G and J may match the habits of the training users better. Nevertheless, even with unseen users, the \name score reflects the system's confidence in sequence selection, which can alert downstream tasks about low-quality inputs.



\subsubsection{Downstream task: respiration rate estimation}
As the last experiment, we evaluate if \name can benefit the performance of the downstream tasks of remote respiration monitoring. We pick respiration rate (RR) calculation as an example. To generate respiration rate ground truth, we first smooth the signal with a Savitzky–Golay filter and then perform peak detection on the smoothed signal. We use the average inter-peak distance to calculate the respiration rate. 

Next, we take the respiration waveform generated using \name and the three baseline methods. We use a signal-processing-based method and a learning-based method and calculate the RR error in beats-per-minute (bpm). The signal processing method is the same as the one introduced above to generate the ground truth. We simply apply the same method to the UWB-measured data. For the learning-based method, we adopt the DeepVS model~\cite{xie2022deepvs}, the same one we used when benchmarking our dataset. The DeepVS model is trained using a training dataset generated with a method similar to our autoregressive model, with $r_0 = 0.5$. 

\begin{table}[h!]
\begin{tabular}{|c|c|c|c|c|}
\hline
Method                                                   & \begin{tabular}[c]{@{}c@{}}MobiVital\\ (Proposed)\end{tabular} & SNR  & CFAR & Variance \\ \hline
\begin{tabular}[c]{@{}c@{}}RR Error\\ (bpm)\end{tabular} & 0.68                                                           & 1.02 & 1.73 & 1.78     \\ \hline
\end{tabular}
\caption{Respiration rate error calculated using signal processing method.}
\label{tab:rate}
\end{table}

\vspace{-20pt}
\begin{figure}[h!]
    \centering
    \includegraphics[width=\linewidth]{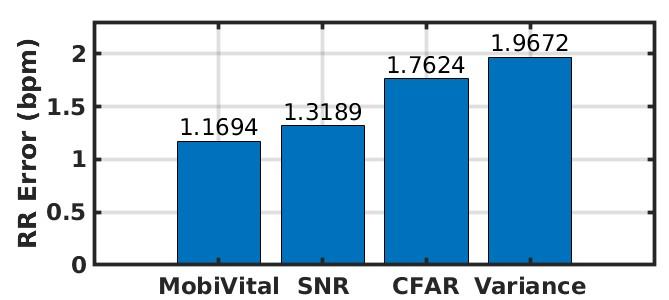}
    \vspace{-10pt}
    \caption{Respiration rate error of the DeepVS model using waveforms extracted from different methods.}
    \label{fig:deepvs}
\end{figure}

Table~\ref{tab:rate} and Figure~\ref{fig:deepvs} show the results of the signal processing method and the DeepVS method, separately. As shown in the figure and the table, \name has a much lower respiration rate error compared to the baseline methods. These results are intuitive, as a higher quality of the measurements naturally leads to better performance in such tasks. Thus, we can see that applying \name to extract respiration sequences with a high correlation with the ground truth is important not only for preserving the waveform but also for improving the accuracy and reliability of downstream analysis.



%% file: Contents/06-RelatedWork.tex
\section{\textbf{Related Work}}\label{Sec:related}

\textbf{UWB Respiration Monitor Systems.} UWB radar has been a popular choice for developing respiration monitoring systems for its low cost, small energy footprints, and relatively high ranging precision~\cite{husaini2022non,zheng2020v2ifi,liu2022vital,yang2019vital,yan2016through, xie2022deepvs, liang2018improved, ren2015noncontact, higashikaturagi2008non}. 
UWB-based respiration monitoring has been used for sleep monitoring~\cite{husaini2022non, kang2020non}, driving monitoring~\cite{zheng2020v2ifi}, and rescuing~\cite{sujatmiko2017review}.  Existing literature has proposed different approaches to measure the respiration rate accurately, where Fourier analysis~\cite{wen2020respiratory, liang2018improved} and wavelet analysis~\cite{liu2022vital, liu2023mmrh} are combined with decomposition algorithms such as empirical mode decomposition (EMD)~\cite{yang2019vital, husaini2022non}, variation mode decomposition (VMD)~\cite{shen2018respiration, wang2023vital}, and singular value decomposition (SVD). Operating at the sub-10 GHz band, UWB radar provides a nice trade-off between penetration and ranging precision, allowing the development of through-wall vital signal monitoring systems~\cite{yan2016through, liang2018improved}. Another advantage of IR-UWB radar lies in its fast-slow-time data structure, which leads to the extracting and separating the respiration signals simultaneously from multiple subjects~\cite{yang2019multi,zeng2020multisense,wang2022capricorn}.

\textbf{Wireless Vital Signal Sensing Platforms.}
Various other sensing technologies have also been researched for non-contact vital sign monitoring. WiFi, mmWave, and RFID-based systems are notable alternatives for detecting physiological motion. WiFi sensing platforms have been heavily researched due to their prevalence, low cost, and ability to penetrate obstacles ~\cite{liu2015tracking,wang2017phasebeat,zhang2023wital,ali2021goodness}. 
Common uses include vital sign monitoring during sleep and daily activities. Liu et al. ~\cite{liu2015tracking} demonstrated tracking respiration and heart rate during sleep using existing WiFi networks. Recently, Zhang et al. proposed Wital~\cite{zhang2023wital}, a vital sign monitoring system capable of tracking multiple targets. These systems typically rely on analyzing channel state information or received signal strength indicators to detect breathing and heartbeat motions. Millimeter-wave (mmWave) technology is also a promising approach for vital sign monitoring due to its high-resolution capabilities~\cite{tang2024bsense, liu2023mmrh, toda2021ecg}. BSense~\cite{tang2024bsense}, developed by Tang et al., is an in-vehicle system leveraging mmWave technology for child detection and monitoring. Deep learning approaches have been applied to reconstruct ECG signals from noisy mmWave signals~\cite{toda2021ecg}. RFID has also been used for contactless vital sign monitoring despite subject's body movement~\cite{liu2019vital}. 

%% file: Contents/07-FinalSessions.tex
\section{Limitations and Future Work}

\noindent\textbf{Dataset Diversity}
The dataset used to develop \name contains 12 users and the duration of the data is about 12 hours for the tripod-mounted platform. Despite being a relatively large user study, the dataset is still limited in terms of diversity. For example, in our evaluations, we partly attribute the inter-user performance variability to not having similar respiration patterns in the training dataset. Also, as a dataset collected in a university, the experiment subjects can have demographic biases such as age and health conditions. Large-scale experiments on big crowds and real respiration data in medical treatments can be a huge booster for systems like \name.

\noindent\textbf{Real-time Breath Coaching System.} Currently, the training and evaluation of \name is conducted offline on a PC with an NVIDIA 1080Ti GPU using a pre-collected dataset. During the development of \name, however, we put out our best effort to provide a lightweight solution. The autoregressive model is simplified to a two-layer LSTM, and the inversion detection algorithm also has a low complexity. Thus, we envision \name can be optimized to execute on mobile platforms such as cell phones. Currently, there are various ``breath coaching'' applications on smartphones and smartwatches, using verbal or visual clues to guide respiration training in yoga or meditation. It would be very beneficiary if we could build a system that can remotely sense your respiration and provide real-time guidance or feedback.

\noindent\textbf{Sensor-subject Relative Motion.}
Motion interference in vital sign monitoring systems is becoming an increasingly interesting topic. Researchers have focused on mitigating the body movements of the subjects~\cite{zhang2021contactless, rong2021motion, gong2021rf}. Recently, researchers have also focused on the movement of the sensor platform~\cite{zhang2022mobi2sense, chang2024msense}. \name mostly considers static subjects during sleeping, sedentary activities, or meditations. The relative motion between the sensor and the subject contaminates the data and may even mask the vital signals. A part of our dataset is collected with a handheld sensor platform. We hope future research efforts can lead us to high-quality respiration waveforms when the user is exercising, or using robot-mounted/ cellphone-mounted sensors.

\section{Conclusions}

In this paper, we pointed out the signal quality issues in UWB-based respiration monitoring, such as deformation and inversion. These issues are largely overlooked in previous works as the waveform quality was not a primary focus.  We presented the \name system, which leverages a self-supervised autoregressive model and a bio-informed algorithm to generate high-quality respiration waveforms from the UWB radar measurement. We demonstrated that our \name score can be a good surrogate for signal quality. The design principle of \name is to harvest the limited generalization ability of the neural networks. While poor performances on out-of-distribution data are generally considered a limitation, sometimes we can inversely leverage this characteristic to benefit engineering designs. In the future, we hope the dataset and algorithms proposed in \name can benefit the design of mobile respiration sensing systems, such as providing more reliable feedback for respiration training systems. 

\begin{acks}
The research reported in this paper was sponsored in part by the Army Research Laboratory (ARL) under Cooperative Agreement W911NF1720196, and the NIH mHealth Center for Discovery, Optimization and Translation of Temporally-Precise Interventions (mDOT) under award 1P41EB028242. The views and conclusions contained in this document are those of the authors and should not be interpreted as representing the official policies, either expressed or implied, of the funding agencies.\looseness=-1
\end{acks}